\documentclass[12pt]{iopart}
\usepackage[utf8]{inputenc}
\usepackage[english]{babel}
\usepackage{graphicx}
\usepackage[colorlinks,linkcolor=blue,citecolor=blue]{hyperref}
\usepackage{subfigure}

\newcommand{\vect}[1]{\bi{#1}}
\newcommand{\matr}[1]{\underline{\underline{#1}}}
\newcommand{\argu}[1]{\!\left(#1\right)}
\newcommand{\set}[1]{\left\lbrace #1 \right\rbrace}

\begin{document}

\title{Magnetic correlations beyond the Heisenberg model in an Fe monolayer on Rh(001)}
\author{A De\'ak$^1$, K Palot\'as$^1$, L Szunyogh$^{1,2}$ and I A Szab\'o$^3$}
\address{$^1$ Department of Theoretical Physics, Budapest University of Technology and Economics, Budafoki \'ut 8., HU-1111 Budapest, Hungary}
\address{$^2$ MTA-BME Condensed Matter Research Group, Budapest University of Technology and Economics, Budafoki \'ut 8., HU-1111 Budapest, Hungary}
\address{$^3$ Department of Solid State Physics, University of Debrecen, Bem t\'er 18/b, HU-4026 Debrecen, Hungary}
\ead{adeak@phy.bme.hu}
\date{}

\begin{abstract}
Motivated by a recent experimental observation of a complex magnetic structure [Takada \emph{et al.}\ 2013 \emph{J.\ Magn.\ Magn.\ Mater.}\ \textbf{329} 95] we present a theoretical study of the magnetic structure of an Fe monolayer deposited on Rh(001). We  use a classical spin Hamiltonian with parameters obtained from \emph{ab initio} calculations  and go beyond the usual anisotropic Heisenberg model by including isotropic biquadratic interactions. Zero-temperature Landau--Lifshitz--Gilbert spin dynamics simulations lead to a complex collinear spin configuration that, however, contradicts experimental finding. We thus conclude that higher order multi-spin interactions are likely needed to account for the magnetic ordering of the system.
\end{abstract}

\section{Introduction}
Magnetism on the nanoscale has become a central aspect of modern technology, with nanostructures being employed in a multitude of industrial applications. For the ongoing development of technology and our basic understanding of the behaviour of novel materials experimental and theoretical studies have to be used in concert. 
Advancements in experimental techniques, in particular, the development of spin-polarized scanning tunneling microscopy (SP-STM) \cite{wiesendanger-2009}, has allowed the direct observation of magnetic structure of various nanosystems at the atomic scale.

 \emph{Ab initio} calculations can greatly help understand the underlying physics behind observed phenomena, and can hint towards hitherto unnoticed features or direct attention to specific systems.
 By growing Fe thin films epitaxially on various transition metal surfaces the effective Fe lattice constant and the hybridization with the substrate can be tuned, providing deep insight into the relationship between electronic and magnetic structure \cite{hardrat-2009, simon-2014}. Using Ir or Rh as substrate \cite{martin-2007, kudrnovsky-2009, deak-2011, spisak-2006, al-zubi-2011, takada-2013} is especially interesting, as their in-plane lattice constant is in between that of bcc-Fe and fcc-Fe.
 
 Recent experimental findings by Takada \emph{et al.}\ \cite{takada-2013} employing SP-STM measurements show that the ground state spin configuration of the Fe$_1$/Rh(001) thin film is a complex noncollinear structure with a $4\times 3$ magnetic unit cell. 
A closer inspection of the spin structure proposed by Takada \emph{et al.}\ \cite{takada-2013} reveals that it is not a conventional spin spiral, since the spins follow a fanning path rather than a spiral along crystallographic directions.

Earlier theoretical investigations of the same system were based on total energy calculations, and considered only a few simple magnetic configurations \cite{spisak-2006,al-zubi-2011}.
\emph{Ab initio} methods demand huge computational effort and many aspects of noncollinear magnetism as well as finite temperature behaviour are too difficult to manage. By matching first principles calculations to classical spin models we may give a less fundamental, yet more tractable and more flexible description of magnetic systems.

In this work we investigate the Fe$_1$/Rh(001) thin film in terms of a spin model consisting of tensorial Heisenberg couplings, as well as higher order two-spin interactions, namely isotropic biquadratic terms. This research is a natural follow-up to our earlier study of the closely related Fe/Ir(001) thin films \cite{deak-2011}.
We use the spin-cluster expansion (SCE) combined with the relativistic disordered local
moment (RDLM) theory \cite{szunyogh-2011} to obtain model parameters. 
We also study the effect of layer relaxations on the magnetic interactions.  The theoretical spin configuration is derived by solving the Landau--Lifshitz--Gilbert equations at zero temperature. The highly collinear 
spin structure obtained due to very strong biquadratic interactions allows us to conclude that higher order multi-spin interactions are needed to describe properly the magnetism of this system.

\section{Methods and computational details}
Computations to obtain the magnetic structure of itinerant systems from first principles usually involve the adiabatic approximation, assuming a separation of time scales between fast (electronic) and slow (spin) degrees of freedom. In terms of the rigid spin approximation the orientational state of the spin system is specified by a set $\set{\vect e}$ of unit vectors $\vect e_i$ describing the orientation of the magnetization at site $i$. 

We study magnetic thin film systems in terms of an extended tensorial Heisenberg model of the form
\begin{eqnarray}
\mathcal H\argu{\set{\vect e}}=\sum\limits_i \vect e_i \matr K_i \vect e_i -\frac{1}{2}\sum\limits_{i\ne j} \vect e_i \matr J_{ij} \vect e_j -\frac{1}{2}\sum\limits_{i\ne j} B_{ij} \left(\vect e_i\cdot \vect e_j\right)^2.\label{eq:spin_model}
\end{eqnarray}
The traceless symmetric matrices $\matr K_i$ describe the on-site anisotropy energy, whereas the exchange tensors  $\matr J_{ij}$ can be decomposed as
\begin{eqnarray}
\matr J_{ij}= J_{ij}^\mathrm{I} \matr I + \matr J_{ij}^\mathrm{A} + \matr J_{ij}^\mathrm{S} \, ,
\end{eqnarray}
where $\matr I$ denotes the unit matrix. These contributions can be easily interpreted, since $J_{ij}^\mathrm{I}$ is the isotropic Heisenberg interaction (note that according to \Eref{eq:spin_model} a positive Heisenberg coupling is ferromagnetic, i.e., it favours parallel alignment of the interacting spins), the antisymmetric matrix $\matr J_{ij}^\mathrm{A}=\frac{1}{2}\left(\matr J_{ij} - \matr J_{ij}^T\right)$ corresponds to the Dzyaloshinskii--Moriya (DM) interaction \cite{dzyaloshinskii-1958,moriya-1960}, and $\matr J_{ij}^\mathrm{S}=\frac{1}{2} \left(\matr J_{ij}+\matr J_{ij}^T- \frac{2}{3} \left(\Tr \matr J_{ij}\right)\cdot \matr I\right)$ describes the second order two-site anisotropy. The last term in \Eref{eq:spin_model} describes the isotropic biquadratic interaction, with a positive coupling preferring collinear (either parallel or antiparallel) orientation.

We use the spin-cluster expansion (SCE) introduced originally by Drautz and F\"ahnle \cite{drautz-2004,drautz-2005} combined with the relativistic disordered local moment (RDLM) method \cite{gyorffy-1985,staunton-2004,staunton-2006} implemented within the framework of the screened Korringa--Kohn--Rostoker (SKKR) multiple scattering theory \cite{zabloudil-2005} to obtain parameters for the spin model from first principles. The SCE allows one to systematically expand the adiabatic magnetic energy surface of a spin system over a set of basis functions constructed from spherical harmonics, leading to a generalized spin model of the form \cite{drautz-2004,drautz-2005}
\begin{eqnarray}
\fl \mathcal H\argu{\set{\vect e}} = \mathcal H_0 + \sum\limits_i \sum\limits_{L\ne \left(0,0\right)} \widetilde J_i^L Y_L\argu{\vect e_i}\nonumber\\
+\frac{1}{2}\sum\limits_{i\ne j} \sum\limits_{L\ne \left(0,0\right)} \sum\limits_{L^\prime \ne \left(0,0\right)} \widetilde J_{ij}^{LL^\prime} Y_L\argu{\vect e_i} Y_{L^\prime}\argu{\vect e_j}+\ldots,\label{eq:sce_spin_model}
\end{eqnarray}
where $Y_L$ are real spherical harmonics and the summations exclude the constant spherical harmonic of composite quantum number $L=\left(\ell,m\right)=\left(0,0\right)$.

On the one hand, the coefficients of the SCE spin model in \Eref{eq:sce_spin_model} can be related to those of the conventional spin Hamiltonian in \Eref{eq:spin_model}. For instance, the $\ell,\ell^\prime=1$ components of $\widetilde J_{ij}^{LL^\prime}$ can be directly related to the $\matr J_{ij}$ tensor, and the isotropic biquadratic coupling can be expressed as \cite{deak-2011}
\begin{eqnarray}
B_{ij}&=-\frac{3}{8\pi}\sum\limits_{m=-2}^2 \widetilde J_{ij}^{(2,m)(2,m)}.
\end{eqnarray}
On the other hand, orthonormality of the spherical harmonics implies that the function defined by \Eref{eq:sce_spin_model} can be projected to obtain the SCE coefficients, e.g.
\begin{eqnarray}
\widetilde J_{ij}^{LL^\prime}=\int \mathrm d^2 e_i \int \mathrm d^2 e_j \left\langle \mathcal H\right\rangle_{\!\vect e_i ,\vect e_j} Y_L\argu{\vect e_i} Y_{L^\prime}\argu{\vect e_j},
\end{eqnarray}
where $\left\langle \mathcal H\right\rangle_{\!\vect e_i ,\vect e_j}$ denotes the partial average of $\mathcal H\argu{\set{\vect e}}$ with fixed spin directions at site $i$ and $j$. Associating these averages of the spin model with the partially averaged \emph{ab initio} grand potential allows us to extract interaction parameters from the electronic structure. The SCE method is especially useful combined with RDLM theory, wherein partial averages of the grand potential can be directly calculated as opposed to using a huge number of individual electronic structure calculations to obtain the averages \cite{deak-2011,szunyogh-2011}. While the method can be formulated to include multi-spin interactions with ease, the computation of such terms is currently beyond the capabilities of our computer code. 

Once a set of model parameters is obtained, these may be used to determine the ground state magnetic configuration. To assess the magnetic ordering we use zero-temperature Landau--Lifshitz--Gilbert spin dynamics simulations as a means for energy minimization. The various terms of the spin model can be easily manipulated in these simulations, in particular to examine how the biquadratic couplings affect the ground state spin structure.

In our calculations the geometry consisted of a single layer of Fe on a semi-infinite Rh fcc-(001) substrate with a semi-infinite vacuum region on top, with an in-plane lattice constant of $2.6898$ \AA\ corresponding to the experimental value for bulk Rh. Layer relaxations were taken into account by varying the distance of the Fe layer from the substrate.
The local spin density approximation (LSDA) of density functional theory was used according to the parametrization of Vosko, Wilk and Nusair \cite{vosko-1980}. The atomic sphere approximation was employed with an angular momentum cutoff of $\ell_\mathrm{max}=2$. For the computation of interaction parameters Brillouin zone (BZ) integrations were carried out with up to 2485 points in the irreducible wedge of the BZ to ensure numerical precision. In spin dynamics simulations the initial state was random, and lattices consisting of $64\times 64$ sites were used with free boundary conditions, or $128\times 128$ in cases where the wavelength of the expected ground state (based on the Heisenberg interaction) was comparable to the smaller system size.

\section{Results}
\subsection{Spin model parameters}
We performed self-consistent-field calculations for the Fe monolayer on Rh(001) for several values of the (inward) layer relaxation of the Fe layer between $0\%$ and $-15\%$, $-9\%$ being the experimental value \cite{begley-1993}. The Fe local spin moment changes moderately from 2.93 $\mu_\mathrm{B}$ to 2.84 $\mu_\mathrm{B}$ in this relaxation range, with 2.87 $\mu_\mathrm{B}$ for the experimental geometry. We found that in every case the on-site anisotropy is much smaller than the two-site contribution, and that the Fe-Fe interaction parameters show strong dependence on the layer relaxation, similar to our previous findings in Fe$_1$/Ir(001) \cite{deak-2011}.
Moreover, due to the smaller atomic number of Rh with respect to Ir, the relativistic interaction terms in the present system are smaller relative to the isotropic Heisenberg couplings than in the case of Ir substrate. 

For a quantitative comparison of energy scales, the dominant value of each component of the Heisenberg tensor is collected in Table \ref{tab:coupling_magnitudes} for a few Fe layer relaxations. For $0\%$, $-5\%$ and $-9\%$ relaxation the components of the first nearest neighbour (NN), while for $-15\%$ relaxation those of the second NN exchange tensor are shown, with the exception of the DM interaction magnitudes which are maximal for second NN's in every case. The isotropic Heisenberg interaction clearly dominates among the bilinear terms, while by increasing the inward relaxation the DM interaction significantly increases. It should be noted, however, that while at smaller layer relaxations the isotropic terms are much larger than the DM interaction, for the experimental geometry and beyond there is only a factor of around 7 between the two contributions, implying that relativistic interactions have to be taken into account for a proper description of the system.

\begin{table} [ht]
\caption{\label{tab:coupling_magnitudes}Comparison of the dominant energy scales of the bilinear interactions in Fe$_1$/Rh(001) for a few values of the Fe layer relaxation, obtained with the SCE method in a DLM reference state. $J^\mathrm{I}$, $D$ and $J_{zz}-J_{xx}$ stand for the isotropic Heisenberg interaction, the magnitude of the DM vector and the two-site anisotropy, respectively. Note the difference in units used.}
\begin{indented}\lineup
\item[]\begin{tabular}{@{}clll}
\br
relaxation & $J^\mathrm{I}$ [mRy] & $D$ [$\mu$Ry] & $J_{zz}-J_{xx}$ [$\mu$Ry]\\
\mr
$0\%$   & 1.90    & 36.8 & 8.73\\ 
$-5\%$  & 1.17    & 63.8 & 8.04\\ 
$-9\%$  & 0.591   & 77.6 & 7.76\\ 
$-15\%$ & \-0.555 & 79.4 & 6.44\\ 
\br
\end{tabular}

\end{indented}
\end{table}

\begin{figure}[!ht]
\includegraphics[width=.5\linewidth]{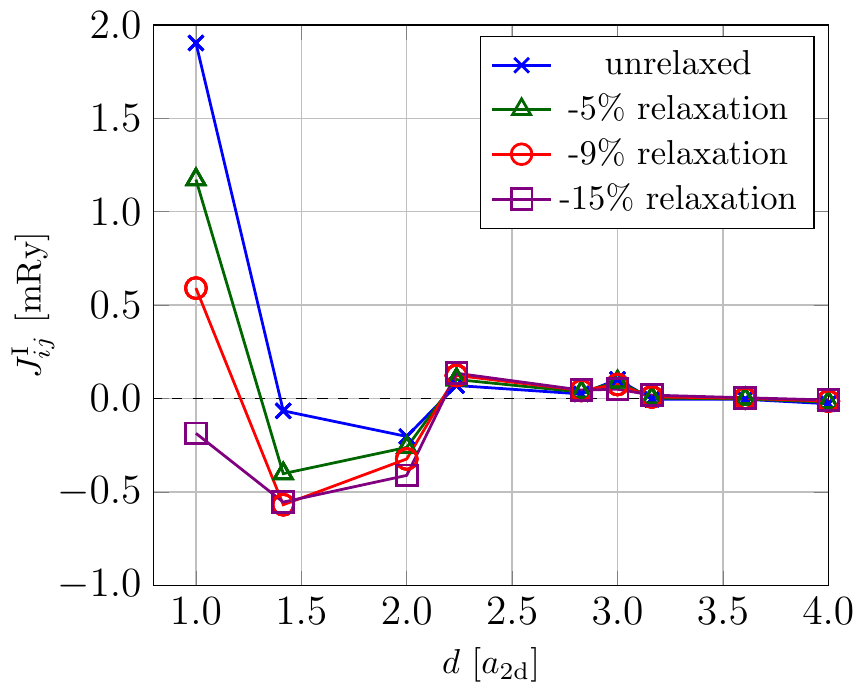}
\includegraphics[width=.5\linewidth]{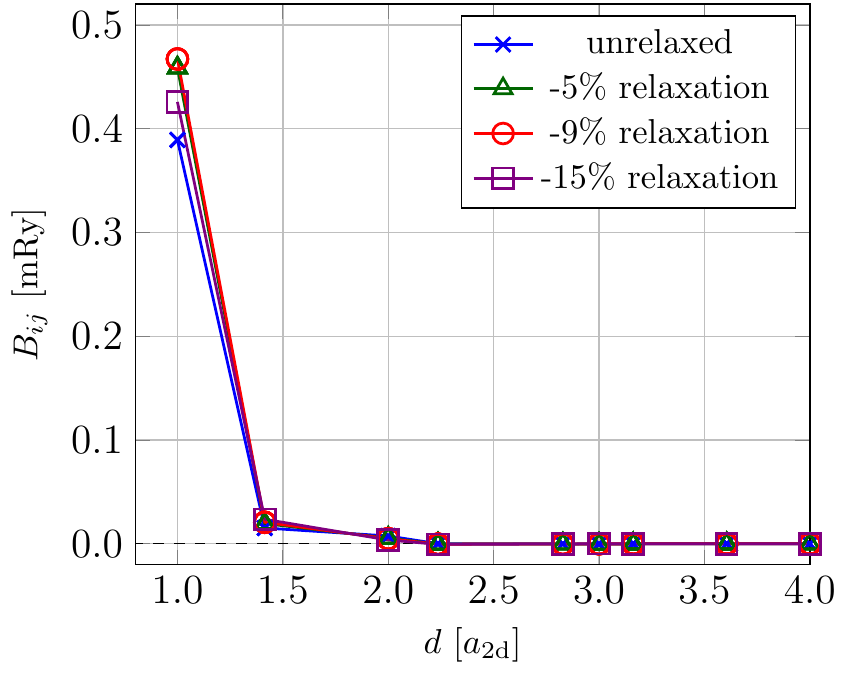}
\caption{\label{fig:jiso_vs_bij}Calculated isotropic Heisenberg ($J_{ij}^{\mathrm{I}}$) and biquadratic ($B_{ij}$) couplings in geometrically unrelaxed and relaxed Fe$_1$/Rh(001) thin films versus interatomic distance ($d$) in units of the in-plane lattice constant ($a_\mathrm{2d}$).}
\end{figure}

In Figure \ref{fig:jiso_vs_bij} the isotropic Heisenberg interaction, $J_{ij}^\mathrm{I}$, and the isotropic biquadratic interaction, $B_{ij}$, are plotted versus the interatomic distance for various values of the Fe layer relaxation. 
For ideal (i.e.\ unrelaxed) geometry there is a strong ferromagnetic (FM) Heisenberg coupling between first NN's, with weakly antiferromagnetic (AFM) second and third NN couplings. 
However, as the Fe inward relaxation is increased, an AFM tendency arises in the Heisenberg interactions. The first  NN couplings decrease in value and become AFM at high relaxations, while the second and third NN AFM couplings become gradually stronger. In particular at the experimental geometry ($-9\%$ Fe relaxation) we can see a weakened first NN FM coupling competing with second and third NN AFM couplings of the same magnitude, leading to strong frustration of the Heisenberg terms. The same qualitative behaviour was also found in Fe$_1$/Ir(001) \cite{deak-2011}, but the AFM tendency is weaker in the present system.

The biquadratic interactions are very strongly localized, only the first nearest neighbour couplings are significant. The first NN values are furthermore only weakly sensitive to the layer relaxation, and their positive value implies that they prefer a collinear arrangement of interacting spins. 
For small relaxations the Heisenberg terms most likely prefer FM ordering of the spins, which also complies with the weaker biquadratic terms. 
However, due to the comparative insensitivity of the biquadratic terms to relaxation, the bilinear and biquadratic couplings are of similar strength near the experimental relaxation. Considering that the spin spiral states preferred by the frustrated Heisenberg terms are incompatible with the biquadratic couplings of positive sign, we anticipate a very strong influence of biquadratic terms on the ground state spin structure due to this additional frustration, to be verified by spin dynamics simulations.

\subsection{Spin dynamics simulations}
For ideal geometry the ground state turned out to be indeed FM, with no frustration between the exchange interactions and the biquadratic terms. As the Fe layer relaxation is increased, however, the competition between these two types of interactions and the frustration of the Heisenberg couplings becomes pronounced, and the ground state is no longer FM.
The change from FM ground state appears quite suddenly. At $-7\%$ relaxation the ground state is still FM, while at $-8\%$ the spin structure becomes a seemingly random collinear ensemble of up and down spins, indicating that the biquadratic coupling dominates the interaction landscape. Figure \ref{fig:jiso_vs_bij} suggests that since the magnitude of the biquadratic terms is weakly sensitive to Fe layer relaxation, this rapid onset of a new type of ground state is due to the decrease of the first NN FM and the increase of second and third NN AFM Heisenberg couplings, leading to frustration of the exchange interactions.

\begin{figure}[!ht]
\begin{center}
\includegraphics[width=.4\linewidth]{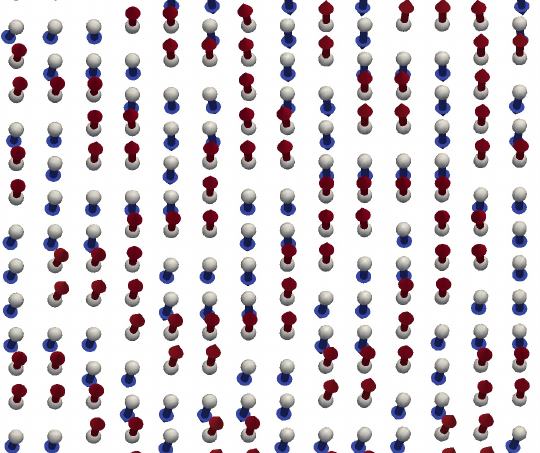}
\caption{\label{fig:expt-sd_sce}Approximate ground state spin configuration simulated with spin-model parameters calculated for experimental geometry. Spins are coloured according to $z$ component, with red colour pointing upward and blue colour pointing downward.}
\end{center}
\end{figure}

This qualitative picture holds true for larger relaxations including the experimental geometry. The exchange interactions by themselves would prefer some kind of spin spiral due to competing FM and AFM couplings (see Figure \ref{fig:jiso_vs_bij}). This frustration allows the comparatively strong first NN biquadratic couplings to overcome the bilinear terms, leading to the collinear spin structure shown in Figure \ref{fig:expt-sd_sce} for experimental geometry, as the biquadratic couplings with positive sign prefer a collinear, either parallel or antiparallel, orientation of interacting spins.

The biquadratic terms themselves do not distinguish between parallel and antiparallel pairs of spins. Omitting bilinear terms, the collinear ground state would be massively degenerate even for a given common axis of orientation, with spins pointing randomly along the common axis. Even though the Heisenberg terms cannot destroy the collinear spin structure preferred by the biquadratic terms, they can lift the aforementioned degeneracy, appearing as a periodic modulation in the random collinear state of spins.

\begin{figure}[!ht]
\begin{center}
\subfigure[]{
\includegraphics[width=.45\linewidth]{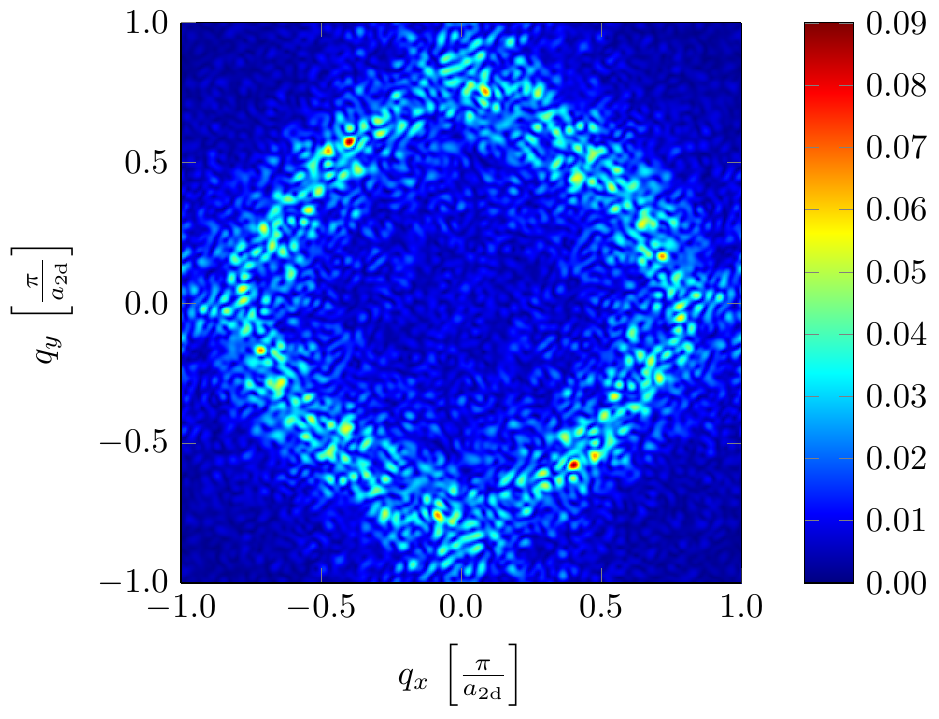}
\label{fig:expt-fourier_sce}
}
\subfigure[]{
\includegraphics[width=.45\linewidth]{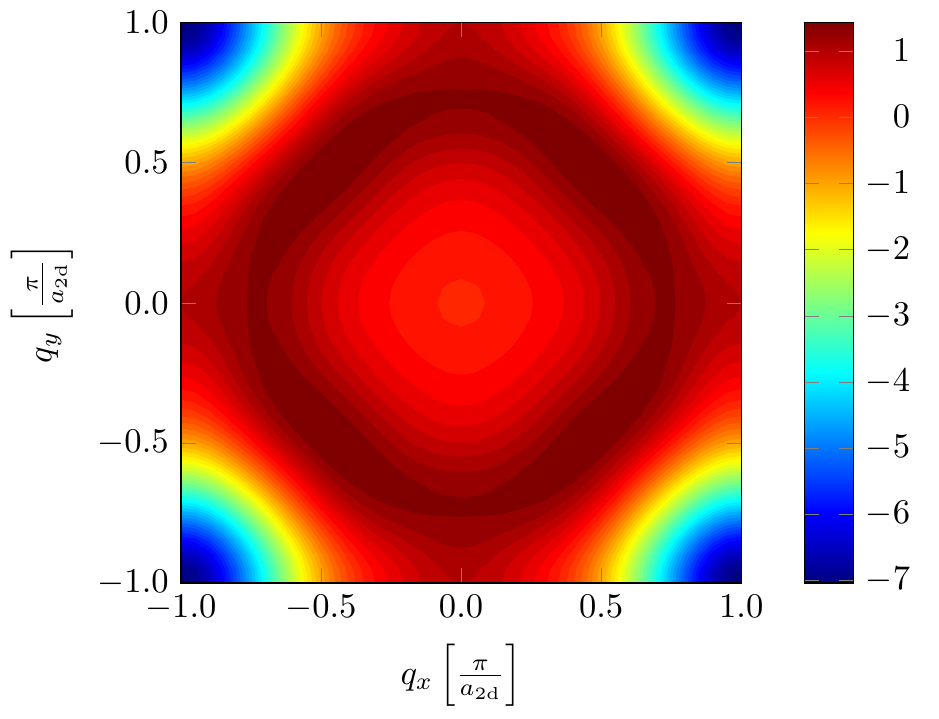}
\label{fig:expt-jq_sce}
}
\caption{\label{fig:expt-four-jq}(a) The scalar lattice Fourier transform, $m\argu{\vect q}$, of the obtained 
spin structure for experimental geometry. (b) The maximal eigenvalues, $J\argu{\vect q}$,  of the lattice Fourier transform of the calculated exchange tensors for experimental geometry in mRy units.}
\end{center}
\end{figure}

To trace this effect we computed the lattice Fourier transform of the spin structures obtained from simulation, 
\begin{eqnarray}
\vect m\argu{\vect q}&=\frac{1}{N}\sum\limits_{j=1}^N \rme^{-\rmi \vect q \vect R_j} \vect{e}_j,
\end{eqnarray}
where $N$ is the number of spins in the sample and $\vect{e}_j$ is the unit vector describing the direction of the spin at site $\vect R_j$. 
The scalar quantity $m\argu{\vect q}=\sqrt{\vect m\argu{\vect q}^\ast \cdot \vect m\argu{\vect q}}$ then plays the role of an indicator of any modulation with wave vector $\vect q$ in the simulation. In Figure \ref{fig:expt-fourier_sce} this scalar is plotted in the entire BZ for experimental geometry. While the value of $m\argu{\vect q}$ is overall very small in the sample indicating that there are only weak correlations in the structure, there is a pattern emerging from the near-zero background in the shape of a rotated square. This is the additional modulation arising from the Heisenberg interaction buried in the spin configuration.

To link the pattern seen in Figure \ref{fig:expt-fourier_sce} to the bilinear couplings we determined the spatial modulation preferred by them. To this end, we used a mean field estimate based on the $\matr \chi\argu{\vect q}$ paramagnetic spin susceptibility, given by
\begin{eqnarray}
\matr \chi\argu{\vect q}=\left[3 k_B T\matr I -\matr J\argu{\vect q}\right]^{-1},
\end{eqnarray}
where $T$ is the temperature, $k_B$ is the Boltzmann constant and $\matr J\argu{\vect q}$ is the lattice Fourier transform of the exchange tensor \cite{deak-2011}. This formula implies that the highest temperature where the susceptibility is singular and, thus, a mean field estimate for the wave vector of the ordered state to which the paramagnetic state is unstable, is given by the global maximum of the eigenvalues of $\matr J\argu{\vect q}$ in the BZ. The maximal eigenvalues, $J\argu{\vect q}$, of the $\matr J\argu{\vect q}$ matrices for experimental geometry are plotted in Figure \ref{fig:expt-jq_sce}.

The similarity between the two quantities shown in Figure \ref{fig:expt-four-jq} is evident. The $J\argu{\vect q}$ shows a shallow, nearly degenerate maximum line along a rounded circle, indicating that the ground state preferred by the Heisenberg terms is somewhere along these $q$ points. It is clear that this maximum line fits perfectly to the Fourier transform of the spin stucture in Figure \ref{fig:expt-fourier_sce}, verifying that the random collinear configuration forced by the biquadratic terms is further modulated by the bilinear interactions.
The qualitative picture seen in Figure \ref{fig:expt-four-jq} is the same for every relaxation equal to or larger than $-8\%$. The sensitivity of the Heisenberg terms to layer relaxation causes the $J\argu{\vect q}$ surfaces to evolve somewhat with increasing inward relaxation, leading to a change in the weak modulation of the $m\argu{\vect q}$ pattern. 

We also performed a set of spin dynamics simulations with the biquadratic terms artificially turned off. As expected, for the larger layer relaxation values the simulations evolved into clear single-$q$ helical spin spirals propagating along the $(1,1)$ direction, with wave vectors agreeing neatly with the numerical maxima of the corresponding $J\argu{\vect q}$ surfaces, being $\vect q=\left(0.47,0.47\right)\frac{\pi}{a_\mathrm{2d}}$ for experimental geometry. The frustration arising from competing first and second NN Heisenberg interactions is overall quite similar to what we found for the Fe$_1$/Ir(001) system \cite{deak-2011}.

\section{Discussion and conclusions}
In summary, we found that the (bilinear) exchange interactions in Fe$_1$/Rh(001) depend strongly on the distance between the Fe overlayer and the substrate, and that their competition leads to strong frustration near the experimental geometry, similarly to what was found in Fe$_1$/Ir(001) \cite{kudrnovsky-2009,deak-2011}. More importantly, we found that biquadratic couplings are comparable to the Heisenberg terms, and even dominate those near the experimental geometry. Consequently, for Fe layer relaxations between $-8\%$ and $-15\%$ we obtained a complex collinear ground state spin structure from spin-dynamics simulations. The fact that the `correction' of the biquadratic terms to the second-order spin model drastically alters the ground state spin configuration implies that the usual anisotropic Heisenberg model is insufficient to describe this system, and one should not even expect to reproduce experimental findings using this approximation.

It is known that multi-spin interactions in small magnetic clusters can be at an energy scale comparable with the bilinear couplings \cite{antal-2007}. Their inclusion into the mapping of \emph{ab initio} total energy might significantly affect the computational estimates for Heisenberg interactions \cite{lounis-2010}. If strong enough, these terms can even influence the ground state spin configuration of magnetic monolayers \cite{hardrat-2009, kurz-2001}, in extreme cases leading to the formation of exotic noncollinear spin structures such as nanoskyrmion lattices \cite{heinze-2011}. 

From the point of view of the SCE, a complete second-order (four-spin) SU(2)-invariant correction to the Heisenberg model is given by a general energy term \cite{drautz-2005}
\begin{eqnarray}
\fl E\argu{2}=\sum\limits_{ij} J^{(2)}_{ij}\left(\vect e_i\cdot \vect e_j\right)^2+\sum\limits_{ijk} J^{(3)}_{ijk}\left(\vect e_i\cdot \vect e_j\right)\left(\vect e_j\cdot \vect e_k\right)\nonumber\\
+\sum\limits_{ijkl} J^{(4)}_{ijkl}\left(\vect e_i\cdot \vect e_j\right)\left(\vect e_k\cdot \vect e_l\right),\label{eq:multiple-spin}
\end{eqnarray}
where the summations always run over sets of distinct sites. This would suggest that the biquadratic terms we chose to be included in \Eref{eq:spin_model} are of the same importance as the three- and four-spin interactions. However, in the framework of the SCE-RDLM method two-site interactions describe two electron propagations to lowest order, while three- and four-spin interactions need at least three and four propagations, respectively (cf.\ Equations (26) and (27) in Reference \cite{szunyogh-2011}). Due to the decay of the Green's function with distance, this suggests that the biquadratic terms should, in general, be more important than multi-spin interactions, albeit the latter can easily be of the same magnitude as multiple-scattering corrections to the former.

All things considered, the spin structures obtained with our simulations are still very unlikely especially with regard to experiments, suggesting that even with the inclusion of the isotropic biquadratic terms we are far from grasping the true magnetic ground state of the system. As for the spin configuration observed in the experiment, upon closer examination one may realize that the $4\times 3$ period in Reference \cite{takada-2013} does not correspond to a single-$q$ spin spiral. A lattice Fourier transform of the proposed configuration seems to show modulations corresponding to three distinct, symmetry unrelated wave vectors, namely $\vect q=\left(1,0\right)\frac{\pi}{a_\mathrm{2d}}$, $\left(0,\frac{2}{3}\right)\frac{\pi}{a_\mathrm{2d}}$ and $\left(\frac{1}{2},\frac{1}{3}\right)\frac{\pi}{a_\mathrm{2d}}$. This fact is in accordance with our findings, in that it seems improbable to be defined by the ground state of a simple Heisenberg model.

\section{Acknowledgments}
AD wishes to thank Leonid Sandratskii for an eye-opening discussion. Financial support was provided in part by the European Union under FP7 Contract No.\ NMP3-SL-2012-281043 FEMTOSPIN (AD, KP, LS). The work was also supported by the projects T\'AMOP-4.2.4.A/1-11-1-2012-0001 (AD) and T\'AMOP-4.2.2.A-11/1/KONV-2012-0036 (LS, IAS) co-financed by the European Union and the European Social Fund.

\end{document}